\documentclass[12pt,epsfig,citesort]{article}
\usepackage{geometry}
\newcommand{\ice}[1]{\relax}
\usepackage{graphicx}
\usepackage{epsfig}
\usepackage{latexsym}
\usepackage{amsmath}
\usepackage{amsfonts}
\usepackage{amsxtra}
\def\be{\begin{equation}}
\def\ee{\end{equation}}
\def\bea{\begin{eqnarray}}
\def\eea{\end{eqnarray}}

\def\ap#1#2#3   {{ Ann. Phys. (NY)} {\bf#1} (#2) #3.}
\def\apj#1#2#3  {{  Astrophys. J.} {\bf#1} (#2) #3.}
\def\apjl#1#2#3 {{ Astrophys. J. Lett.} {\bf#1} (#2) #3.}
\def\app#1#2#3  {{ Acta. Phys. Pol.} {\bf#1} (#2) #3.}
\def\ar#1#2#3   {{ Ann. Rev. Nucl. Part. Sci.} {\bf#1} (#2) #3.}
\def\cpc#1#2#3  {{ Computer Phys. Comm.} {\bf#1} (#2) #3.}
\def\err#1#2#3  {{ Erratum} {\bf#1} (#2) #3.}
\def\ib#1#2#3   {{ ibid.} {\bf#1} (#2) #3.}
\def\jmp#1#2#3  {{ J. Math. Phys.} {\bf#1} (#2) #3.}
\def\ijmp#1#2#3 {{ Int. J. Mod. Phys.} {\bf#1} (#2) #3.}
\def\jetp#1#2#3 {{ JETP Lett.} {\bf#1} (#2) #3.}
\def\jpg#1#2#3  {{ J. Phys. G.} {\bf#1} (#2) #3.}
\def\mpl#1#2#3  {{ Mod. Phys. Lett.} {\bf#1} (#2) #3.}
\def\nat#1#2#3  {{ Nature (London)} {\bf#1} (#2) #3.}
\def\nc#1#2#3   {{ Nuovo Cim.} {\bf#1} (#2) #3.}
\def\nim#1#2#3  {{ Nucl. Instr. Meth.} {\bf#1} (#2) #3.}
\def\np#1#2#3   {{ Nucl. Phys.} {\bf#1} (#2) #3.}
\def\pcps#1#2#3 {{ Proc. Cam. Phil. Soc.} {\bf#1} (#2) #3.}
\def\pl#1#2#3   {{ Phys. Lett.} {\bf#1} (#2) #3.}
\def\prep#1#2#3 {{ Phys. Rep.} {\bf#1} (#2) #3.}
\def\prev#1#2#3 {{ Phys. Rev.} {\bf#1} (#2) #3.}
\def\prl#1#2#3  {{ Phys. Rev. Lett.} {\bf#1} (#2) #3.}
\def\prs#1#2#3  {{ Proc. Roy. Soc.} {\bf#1} (#2) #3.}
\def\ptp#1#2#3  {{ Prog. Th. Phys.} {\bf#1} (#2) #3.}
\def\ps#1#2#3   {{ Physica Scripta} {\bf#1} (#2) #3.}
\def\rmp#1#2#3  {{ Rev. Mod. Phys.} {\bf#1} (#2) #3.}
\def\rpp#1#2#3  {{ Rep. Prog. Phys.} {\bf#1} (#2) #3.}
\def\sjnp#1#2#3 {{ Sov. J. Nucl. Phys.} {\bf#1} (#2) #3.}
\def\spj#1#2#3  {{ Sov. Phys. JEPT} {\bf#1} (#2) #3.}
\def\spu#1#2#3  {{ Sov. Phys.-Usp.} {\bf#1} (#2) #3.}
\def\zp#1#2#3   {{ Zeit. Phys.} {\bf#1} (#2) #3.}

\def\beq{\begin{equation}}
\def\eeq{\end{equation}}
\def\bea{\begin{eqnarray}}
\def\eea{\end{eqnarray}}

\begin{document} 
%\date{\today}
\date{} % this for the final version

%%%%%%%%%%%%%%%%%%%%%%%%%%%%%%%%%%%%%%%%%%%%%%
%% title page 
%%%%%%%%%%%%%%%%%%%%%%%%%%%%%%%%%%%%%%%%%%%%%%

\title{\bf \Large WMAP Dark Matter Constraints on Yukawa Unification 
with Massive Neutrinos}
\author{ \bf \normalsize 
M.E. Gomez $^{a}$,
S. Lola $^b$, 
P. Naranjo $^b$ 
and J. Rodriguez-Quintero $^{a}$
%Galeon S.L.
}

\maketitle

\begin{center}
$^{a)}$ Departamento de F\'{\i}sica Aplicada, University of Huelva, 
21071 Huelva, Spain \\
$^{b)}$ Department of Physics, University of Patras, 26500 Patras, Greece
\end{center}

\vspace*{2 cm}

\begin{quote}

\begin{abstract}
We revisit the WMAP dark matter constraints on Yukawa Unification 
in the presence of massive neutrinos. The large lepton mixing indicated 
by the data may modify the predictions for the bottom quark mass,
enabling Yukawa unification also for large $\tan\beta$, and for positive
$\mu$ that was previously disfavoured.
As a result, the allowed parameter space
for neutralino  dark matter increases for positive $\mu$, particularly 
for areas with resonant enhancement of the neutralino relic density.
On the contrary, a negative $\mu$ is not easily compatible with
large lepton mixing and Dirac neutrino Yukawa couplings, and the
WMAP allowed parameter space is in this case strongly constrained.

\end{abstract}

\vspace*{2 cm}

%\begin{flushright}
%{\small UHU-GEM/08-11}\\
%{\small CPHT RR 038.0605}\\
%{\small LPT-Orsay/08-28}\\
%\end{flushright}

%\vfill
%\newpage

\end{quote}
%\end{titlepage}
\pagebreak

%%%%%%%%%%%%%%%%%%%%%%%%%%%%%%%%%%%%%%%%%%%%%%%%%%%%%%%%%
%% end of title page
%%%%%%%%%%%%%%%%%%%%%%%%%%%%%%%%%%%%%%%%%%%%%%%%%%%%%%%%%

%%%%%%%%%%%%%%%%%%%%%%%%%%%%%%%%%%%%%%%%%%%%%%%%%%%%%%%%%
%% body of the paper
%%%%%%%%%%%%%%%%%%%%%%%%%%%%%%%%%%%%%%%%%%%%%%%%%%%%%%%%%

\section{Introduction}
Reconciling the Cold Dark Matter (CDM) predictions of 
supersymmetric models  with
the stringent constraints from the Wilkinson Microwave Anisotropy Probe
(WMAP) has been one of the challenges  within the particle physics community 
in recent years. The amount of CDM  deduced from the WMAP data is 
given by \cite{wmap}
\begin{equation}
\Omega_{CDM} h^2 =0.111^{+0.011}_{-0.015} \;,
\end{equation}
where  $h$ is the Hubble parameter in units of 100km/s/Mpc
and  $\Omega_{CDM} =\rho_{CDM}/\rho_c$ the ratio
of the matter density of cold dark matter, $\rho_{CDM}$, over
the  critical density, $\rho_c$, that leads to a flat Universe.
This puts severe constraints on the type
and parameter space of unified theories with Dark Matter candidates.
One of the most promising frameworks in this respect is provided by
supergravity models that conserve R-parity ~\cite{msugra}, and thus
predict a stable lightest supersymmetric particle (LSP). This particle
has to be neutral, and an obvious candidate is the neutralino ~\cite{goldberg}, 
although gravitinos are also well-motivated \cite{Ellis:2003dn}
\footnote{In fact, gravitinos can be dark matter even in theories 
with R-parity violation, if their lifetime is larger than the 
age of the universe \cite{Rvio-gra}.}.

If one imposes Yukawa unification, as expected from Grand Unified Theories
(GUTs),  the solutions become more predictive and
additional constraints are imposed on the model parameters
\cite{sugradark,profumo}.
The same  holds for bounds from Flavour Changing Neutral Current 
(FCNC) processes, which have to be included in the analysis and strongly 
constrain the allowed supersymmetric parameter space \cite{roberto}.

In addition to the above, the neutrino data of the past years
provided evidence for the existence of neutrino oscillations and masses,
pointing for the first time to physics beyond the Standard Model 
\cite{neu-data}. By now, it has been established that the 
atmospheric and solar mixing angles $\theta _{23}$
and $\theta _{12}$ are large and  that the squared mass 
differences are $\Delta m_{atm}^2\simeq 
2.5\times 10^{-3}$ eV$^2$ and $\Delta m_{sol}^2\simeq 8\times 10^{-5}$ eV$^2$ 
respectively \cite{data-fits}.
As expected, however, the additional interactions required to
generate neutrino masses also affect the energy dependence of the
couplings of the Minimal Supersymmetric Standard Model (MSSM), 
and thus modify the Yukawa unification predictions.
In fact, a first observation had been that the additional interactions of
neutrinos to the tau would spoil bottom-tau Unification for small
values of $\tan\beta$ \cite{VB}, 
where we are away from any fixed-point structure
while the  corrections to the bottom quark mass are small.
Subsequently, however, it has  been realized that the large lepton mixing
could naturally restore unification, and even enable Unification for intermediate
values of $\tan\beta$ that were previously disfavoured \cite{LLR,CELW}. It is interesting to note that for the cosmologically favoured area, it is also 
possible to observe tau flavour violation at the LHC, in the framework of non-minimal supersymmetric Grand Unification  \cite{edson}.

In this work, we revisit the issues of Dark Matter and Yukawa Unification
taking into account the effects of massive neutrinos, and large
lepton mixing, as indicated by the data. As a first step, we extend
previous results to large $\tan\beta$, finding significant effects
on the allowed parameter space and on $m_b$.
In fact, it turns out that Yukawa Unification in the presence
of neutrinos is also compatible with a positive $\mu$ and large 
$\tan\beta$, unlike what
happens if the effects of neutrinos are ignored \cite{GNIS1,GLP,shafi}. 
Passing to the relic density of neutralinos,
we studied the consequences of the above, particularly for  
the $\chi - \tilde{\tau}$ coannihilation region and
for resonances in the $\chi-\chi$ annihilation channel, finding 
once more sizeable 
effects, which are further amplified when large lepton mixing is combined
with quantum corrections above the GUT scale.

\section{Massive Neutrinos and Gauge and Yukawa Unification}

The most straightforward extension of the Standard
Model that can accommodate neutrino masses
is to include three very heavy right-handed 
neutrino states and assume that the smallness of masses
arises from the See-Saw mechanism \cite{seesaw}.
In this case, the predictions for $m_b$ and unification clearly get modified.
In particular,  radiative corrections  from  the neutrino Yukawa couplings
have to be included for renormalization group runs from 
$M_{GUT}$ to $M_{N}$ 
(scale of the heavy right-handed 
neutrinos) \cite{barger}.  Below $M_N$, right-handed neutrinos decouple from the 
spectrum and an effective  see-saw mechanism is operative (with the 
neutrino mass operator running down to low energies). The relevant equations are given in \cite{GL} and are summarized in the Appendix.

In addition, if the GUT scale 
lies significantly below a scale $M_X$, at which gravitational 
effects can no longer be neglected, the 
renormalization of couplings
at scales between $M_X$ and $M_{GUT}$ may induce 
additional effects to the running.
The simplest such example is provided within the framework of
the minimal supersymmetric
SU(5) GUT, and the renormalization group equations (RGEs) in this 
case are given in \cite{Hisano} and are also summarized in the Appendix.
These runs affect Yukawa unification via the supersymmetric 
corrections to the bottom quark mass; however, since in 
the leading-logarithmic approximation the induced
modifications to soft masses are proportional 
to the $V_{CKM}$ mixing \cite{Hisano}, they are significantly suppressed.

Nevertheless, it has been realized  that the influence 
of the runs above the GUT scale on the Dark Matter abundance can be 
very sizeable \cite{Mamb}, due to  changes in
the relation between $m_{\tilde{\tau}}$ and $m_{\chi}$, 
which is crucial in the coannihilation 
area. In dark matter calculations, therefore, the possibility
of such effects has to be also analyzed, and compared to the more
standard scenarios. 

Our starting point will be the $SU(5)$ superpotential 
at  $M_X$ which (omitting generation indices) is given by
\begin{equation}
\label{W}
{\mathcal{W}}_X=T^T\,\lambda_u^{\delta}\,T\,H + 
T^T\,\lambda_d\,\bar{F}\,\bar{H} + \bar{F}^T\,\lambda_N^{\delta}\,S\,H + 
S^T\,M_N\,S
\end{equation}
Here, $T, \bar{F}$ and $S$ are the \textbf{10, 5$^\ast$} and \textbf{1} 
$SU(5)$ superfields, respectively, and  
$H$ and $\bar{H}$ are the \textbf{5} and \textbf{5$^{\ast}$} 
Higgs superfields. 
The couplings 
$\lambda_{u,d,N}$ stand for the up-type quark, 
down-type quarks/charged lepton and Dirac neutrino Yukawa matrices.
The symbol $\delta$ stands for \emph{diagonal} 
(the up- and down-type quark Yukawa matrices 
cannot be diagonalized simultaneously). 
Finally, $M_N$ is the right-handed neutrino mass matrix. 
In addition, we work with the soft SUSY-breaking Lagrangian 
\begin{eqnarray}
\label{Lsoft}
{\mathcal{L}}_{soft} & = & \tilde{T}^{\dagger}\,m_{\mathbf{10}}^2\,\tilde{T} + 
\tilde{\bar{F}}^{\dagger}\,m_{\mathbf{5}}^2\,\tilde{\bar{F}} + 
\tilde{S}^{\dagger}\,m_{\mathbf{1}}^2\,\tilde{S} + \nonumber \\
                    &   & m_h^2\,h^{\dagger}\,h 
+ m_{\bar{h}}^2\,\bar{h}^{\dagger}\,\bar{h} + 
M_5\,\lambda_{5L}\,\lambda_{5L} + \mathrm{h.c.} + \nonumber \\
                    &   & \{\tilde{T}^T\,A_u\,\tilde{T}\,h + 
\tilde{T}^T\,A_d\,\tilde{\bar{F}}\,\bar{h} + 
\tilde{\bar{F}}^T\,A_N\,\tilde{S}\,h + \mathrm{h.c.}\}
\end{eqnarray}
where $m_{\mathbf{10,5,1}}$ are the \textbf{10, 5$^\ast$} and \textbf{1} 
$SU(5)$ superfields masses, respectively, and $m_{h,\bar{h}}$ are the 
\textbf{5} and \textbf{5$^{\ast}$} Higgs superfields masses. $A_{u,d,N}$ are 
the up-type quark, down-type quark/charged lepton and Dirac neutrino 
trilinear terms. Finally, $\lambda_{5L}$ is the $SU(5)$ gaugino 
and $M_5$ its soft Majorana mass. In the mSUGRA scenario, 
universal boundary conditions are assumed at the  gravitational scale
\begin{equation}
m_{\mathbf{10}}^2=m_{\mathbf{5}}^2=m_{\mathbf{1}}^2=
m_h=m_{\bar{h}}=
m_0^2,
\end{equation}
\begin{equation}
A_f=a_0\,\lambda_f,\,\,\,\,\,\,f=\{u,d,\nu\}
\end{equation}

The physical values of the masses will then be obtained by
integrating  the renormalization group equations from $M_X$ 
down to low energies. In the first part of the runs we work
with the RGEs of SU(5) from the 
high energy scale $M_X$ down to $M_{GUT}$.
Subsequently, we run the MSSM RGEs,
supplemented with right-handed neutrinos, from $M_{GUT}$ to
the right-handed  neutrino scale, $M_N$. At this scale,
right-handed neutrinos decouple from 
the spectrum, leaving us with the MSSM RGEs and the see-saw 
operator (which is also renormalized  down to the low energy scale).
The effects of neutrinos on Yukawa unification crucially depend on
the magnitude of the dominant neutrino Dirac Yukawa coupling, $\lambda_N$,
and consequently the magnitude of $M_N$ (which determines
$\lambda_N$ through the See-Saw conditions). Obviously, for low
$M_N$ and $\lambda _N$, the effects are less significant.

Large neutrino Yukawa couplings affect unification by
increasing the predicted value of $m_b(M_Z)$.
This can be understood  for small $\tan\beta$ by 
simple, semi-analytic expressions \cite{LLR,CELW} 
since only the top and the Dirac-type 
neutrino Yukawa couplings ($\lambda_t$ and $\lambda_N$)
may be large at the GUT scale. In this case, 
the RGEs take  simple form
\bea
16\pi^2 \frac{d}{dt} \lambda_t= \left(
    6 \lambda_t^2  + \lambda_N^2
    - G_U\right)  \lambda_t,  
 & & 16\pi^2 \frac{d}{dt} \lambda_N = \left(
    4\lambda_N^2  + 3 \lambda_t^2
   - G_N \right) \lambda_N \nonumber   \\
 16\pi^2 \frac{d}{dt} \lambda_b =
    \left(\lambda_t^2 - G_D \right) \lambda_b,  
 & & 16\pi^2 \frac{d}{dt} \lambda_{\tau} =\left( \lambda_N^2
  - G_E \right) \lambda_{\tau}
 \label{eq:rg4}
 \eea
 Here, $\lambda_\alpha$, $\alpha=U,D,E,N$, represent the
 $3 \times 3$ Yukawa matrices for the up and down quarks, charged
lepton and Dirac neutrinos,
and $G_{\alpha}= \sum_{i=1}^3c_{\alpha}^ig_i(t)^2$ are
functions of the  gauge couplings with the
coefficients $c_{\alpha}^i$'s as in \cite{VB}.
Denoting by $\lambda_{b_0},{\lambda_{\tau_0}}$
the $b$ and $\tau$ couplings
at the  unification scale, it is found that
 \begin{equation}
 \lambda_{b}(t_N)=\rho
 \xi_t\frac{\gamma_D}{\gamma_E}\lambda_{\tau}(t_N), ~~~
\rho=\frac{\lambda_{b_0}}{\lambda_{\tau_0}\xi_N}
 \end{equation}
 where $\gamma_\alpha(t)$ and
 $\xi_{i}$ depend only on
 gauge and Yukawa couplings. Let us then assume successful 
$b-\tau$ unification at $M_{GUT}$, 
with  $\lambda_{\tau_0} =\lambda_{b_0}$. In the absence
 of right-handed neutrinos $\xi_N \equiv 1$, thus
 $\rho =1 $. In the presence of them, however,  $\lambda_{\tau_0}
 =\lambda_{b_0}$ at the GUT scale implies that 
$\rho \neq 1$  (since $\xi_N<1$). To restore $\rho$ to unity, a 
deviation from $b-\tau$ unification would seem to be required.
However,  
large lepton mixing is reconciled with Yukawa unification, by
making the simple observation
that the $b-\tau$ equality at the GUT scale refers to the
$(3,3)$ entries of the charged lepton and
down quark mass matrices.
It is then possible to assume mass textures,
such that, after the diagonalization at the
GUT scale, the $(m^{diag}_E)_{33}$ and $(m^{diag}_D)_{33}$
entries are no-longer equal \cite{LLR}.
The simplest example
the one with symmetric mass matrices:
\begin{equation}
{\mathcal {M}}_d^0 \propto A\,\left(\begin{array}{cc}
y & 0 \\
0 & 1  
\end{array}\right), \,\,\,\,
{\mathcal {M}}_{\ell}^0 \propto A\,\left(\begin{array}{cc}
x^2 & x \\
x & 1 
\end{array}\right)
\label{Ml}
\end{equation}
Here, $A$ may be identified with $m_b(M_{GUT})$. While the 
textures ensure equality of the (3,3) elements of the down and charged 
lepton mass matrices, the eigenvalue of the
charged lepton mass matrix is not 1, but $1+x^2$, thus implying that 
$\lambda_b \neq \lambda_\tau$ after diagonalization.
Within this framework,  the issue of Yukawa Unification in the presence 
of large lepton mixing has been analyzed in \cite{LLR, CELW} for 
small and moderate  values of $\tan\beta$. 

In our current work, we extend these results
to all values of $\tan\beta$ and both signs of $\mu$,  
taking appropriately into account
the large supersymmetric corrections to $m_b$ \cite{Hall,CW,GNIS0},
and using up-to-date experimental bounds. 
Potentially large lepton 
mixing effects will be taken into account by imposing the GUT condition 
$\lambda_\tau=\lambda_b(1+\delta)$  where $\delta$ is determined by requiring 
that $m_b(M_Z)$ is in the allowed experimental range. 
The third generation Dirac neutrino Yukawa coupling  
$\lambda_N$ is determined through the see-saw mechanism
 by assuming 
$m_{\nu_3}=0.05$~eV and a heavy Majorana neutrino scale
$M_N=3 \times 10^{14}$~GeV (a value that ensures that
$\lambda _N$ stays within the perturbative 
regime).  On the other hand, the behaviour associated with the absence 
of a Dirac neutrino coupling should be recovered by appropriately lowering 
$M_N$ and thus
$\lambda _N$. 
The resulting effects to the allowed parameter space for unification are 
significant, and give interesting information on the magnitude and origin of 
lepton mixing. We then proceed to discuss in detail the cosmological 
implications in the framework of the WMAP data.

\section{Yukawa Unification in SUSY Models}

While gauge unification is considered as one of the most attractive 
features of Supersymmetry, the relations among Yukawa couplings 
derived by embedding $SU(3)\times SU(2)\times U(1)$ in a larger gauge
group are  less clear. The charged fermion mass hierarchies
indicate that any mass correlations induced by a unified gauge 
symmetry will be more explicitly manifest in the relation 
between the Yukawa couplings
of the third generation,  $\lambda_t$, $\lambda_b$, $\lambda_\tau$. 
How this works depends on the theoretical framework that 
defines the initial conditions. In $SO(10)$, for instance,
all fermions are in the same representation, thus,
in the simplest realizations, neutrino couplings 
will be unified with the rest. 
In $SU(5)$, the field structure is
$(Q,u^{c},e^{c})_{i} \in {\tt 10}$ of $SU(5)$
and $(L,d^{c})_i  \in { \tt \overline{5}}$, only implying  
$b-\tau$ unification; neutrinos are singlets and
thus there is a freedom of choice (although the fact that they
couple to the same Higgs as the up-type quarks could be a motivation for
some link with the top coupling). 
Clearly, the more restrictive the unification schemes,
the stronger the correlations between quarks and leptons.

In supersymmetric models, unification turns out to be very sensitive to
the model parameters. Among others,
the compatibility of $b-\tau$ unification 
with the observed $b$ and $\tau$ masses 
has a sensitive dependence on the sign of the 
Higgs mixing parameter, $\mu$, and on the
details of the superparticle 
spectrum~\cite{Komine:2001rm}. Moreover, the universality condition  
on the soft terms can be removed as in \cite{profumo},
resulting in an enhancement of the allowed parameter space. In fact, 
it turns out that full  third generation 
Yukawa Unification $\lambda_t=\lambda_b=\lambda_\tau$ in the CMSSM 
fails to accurately predict the experimental values 
of the third generation fermion masses,
unless one goes to a large $\tan\beta$ regime with a heavy
spectrum and a dark matter abundance that would tend to 
overclose the universe.  

To correctly obtain  pole masses within this framework, 
the standard model and supersymmetric  threshold corrections
have to be included; for the bottom quark, these
corrections result to a $\Delta m_b$ that 
can be very large, particularly 
for large values of $\tan\beta$.
In mSUGRA, in the absence of phases, 
$\Delta m_b$ has the sign  of $\mu$ \cite{Hall,CW,GNIS0},
which is required in order to 
obtain a $b$-quark mass in the allowed range in 
models with $b-\tau$ unification. 
However, this also results to
a positive supersymmetric
contribution to BR($b\rightarrow s \gamma$), which can be
compatible with the bounds only for a heavy 
sparticle spectrum. 

The allowed parameter space is nevertheless extremely constrained
from  the bounds on Flavour Changing Neutral Currents, and the 
new bounds on  $b \rightarrow s \gamma$ \cite{Barberio} are crucial for 
the whole discussion and comparisons with the SM prediction \cite{Misiak}. 
The $g_{\mu}-2$ constraints from Brookhaven National Laboratory (BbNL)
\cite{Bennett:2004pv} are also relevant  since
the supersymmetric contribution to 
the muon anomalous magnetic moment
takes the sign of $\mu$ in mSUGRA~\cite{lncn}.
There are also some uncertainties  $\mathcal{O}$($\alpha ^2$) 
in the Standard Model prediction,
due to the hadronic vacuum polarization correction 
~\cite{Miller:2007kk}. 
For a heavy sparticle spectrum
the SUSY contribution to $\Delta a_{\mu}$ 
is small with respect to the Standard Model one. However, we do not take into account 
that constraint derived from $g_\mu-2$, 
since there is no significant deviation from the SM predictions if the 
hadronic contribution is calculated using $\tau$-decay data (the 
current experimental value deviates by 3.4 $\sigma$ from the SM prediction
if the hadronic vacuum polarization is determined using $e^+e^-$ annihilation data).

Let us also summarize a few facts on 
the possible range of the mass of the bottom quark:
The 2-$\sigma$ range for 
the  $\overline{MS}$ bottom running mass,
$m_b(m_b)$, is from  4.1-4.4 GeV 
(with corresponding pole masses from 4.7 to 5 GeV). 
We also know that $\alpha_s(M_Z)=0.1172 \pm 0.002$, 
and the central  value of $\alpha_s$ corresponds 
to $m_b(M_Z)$ from  2.82 to 3.06 GeV,
while the value $m_b(m_b)=4.25$~GeV is mapped to
$m_b(M_Z)=2.92$~GeV. The allowed strip for $m_b(M_Z)$ moves to 
2.74 GeV $<m_b(M_Z)<$ 3.014 GeV  for  $\alpha_s^{max}$ 
and to 2.862 GeV $<m_b(M_Z)<$ 3.114 GeV for  $\alpha_s^{min}$.
%\footnote{We use the LEP2 bounds on the mass of the lightest Higgs boson, 
%namely $m_h>114.4$ GeV. As far as the top mass is concerned, we take the 
%most recent value reported by \cite{TEWG}.}

\begin{figure}[!h]
\hspace*{-0.3 cm}
\includegraphics[width=8cm,height=9cm]{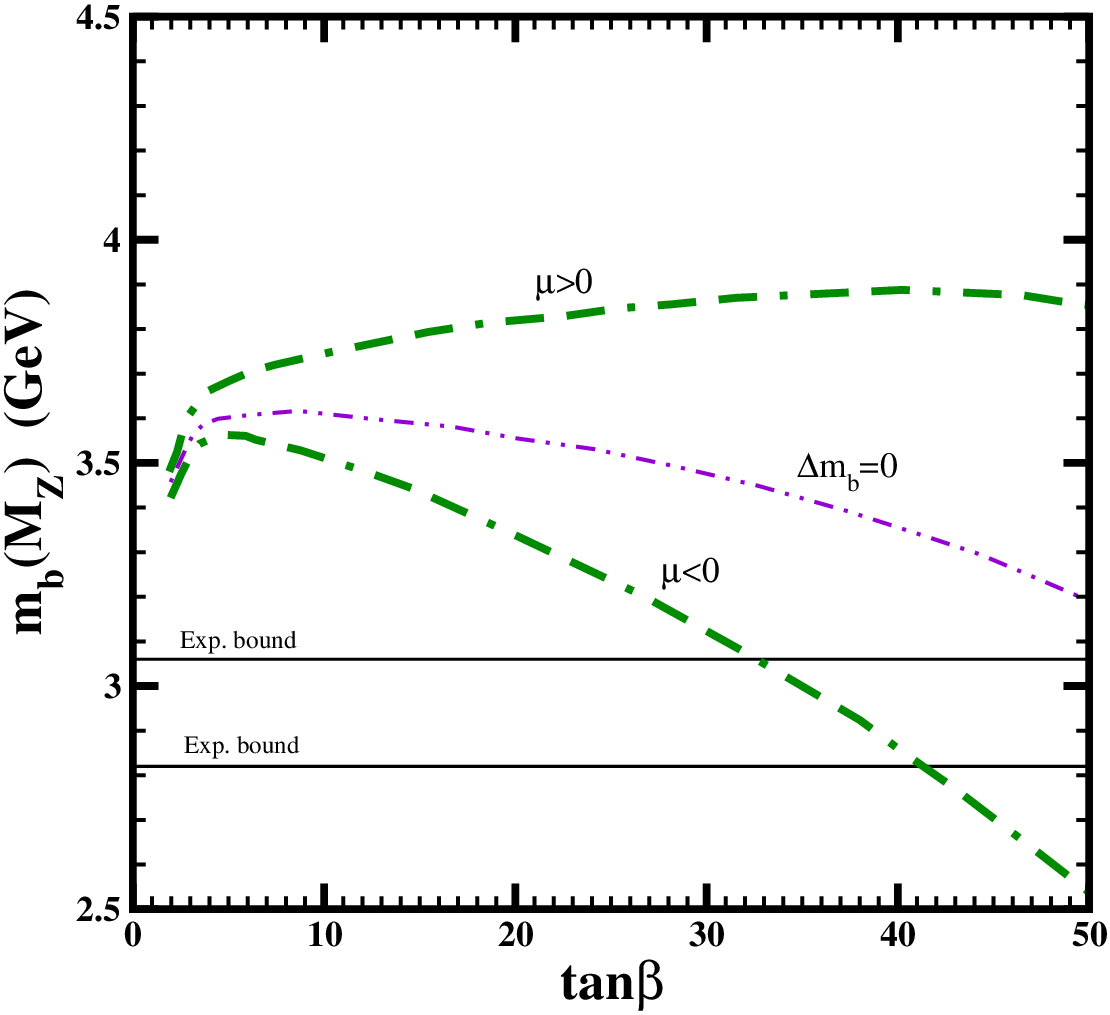}
\includegraphics[width=8cm,height=9cm]{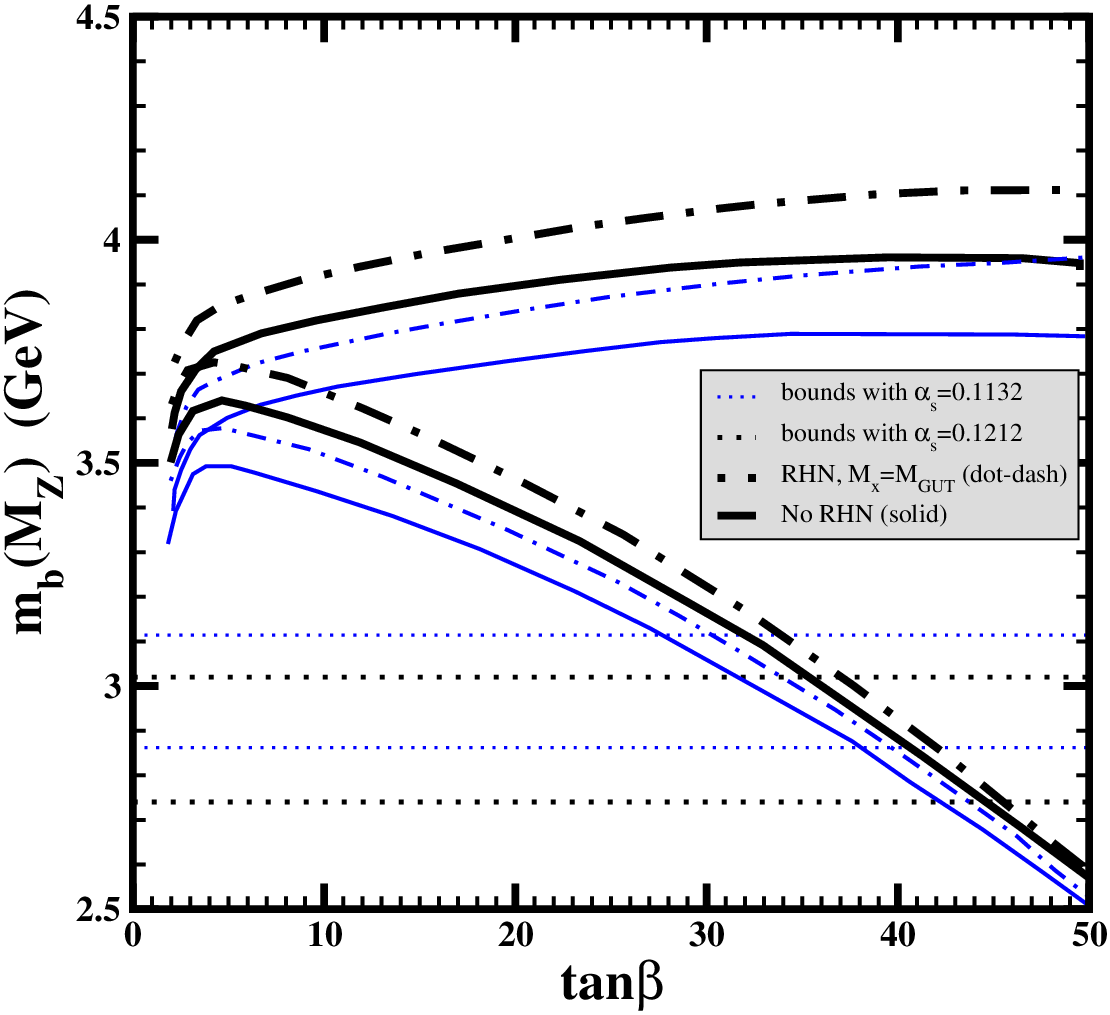}
\caption{ \it 
The value of $m_b(M_Z)$ versus $\tan \beta$ assuming 
$\lambda_b = \lambda_{\tau}$ at the high scale in the absence 
(left plot) and presence (right plot)
of massive neutrinos and for both signs of $\mu$ 
for the following set of parameters:
$M_{1/2}=800$ GeV , $A_0=0$ GeV, $m_{0}=600$ GeV. The experimental range 
of $m_b$ (horizontal lines) shown in the left pannel is computed for the central 
value of $\alpha _s\left(M_Z\right)$. In the right panel, the solid lines 
are obtained within the MSSM, while the dot-dash lines include Dirac 
Yukawa couplings up to the scale $M_N=3\times 10^{14}$ GeV. Here we take the lowest 
(blue thin) and highest (black thick) experimental 
values of $\alpha _s\left(M_Z\right)$.}
\label{mb1}
\end{figure}

In the left panel of Fig. \ref{mb1}, we summarize the predictions for $m_b$
in mSUGRA, where all CP phases are either 
zero or $\pi$ (defining the  sign of $\mu$).  In order to discuss 
the dependence of $m_b(M_Z)$ on $\tan\beta$, 
we consider the following set of soft parameters: 
$M_{1/2}=800$ GeV, $A_0=0$, $m_0=600$ GeV. 
We study this set for both $\mu>0$ and $\mu<0$,
setting $\alpha_s(M_z) = 0.1172$.
We also include a reference line without the SUSY corrections
to the bottom mass (double-dot-dash line, for $\Delta m_b=0$). 
The figure exhibits the known 
fact  that in the absence of phases or
large trilinear terms,  $\Delta m_b$ is positive for $\mu$ positive,
and therefore the theoretical prediction for the $b$ quark pole mass is 
too high to be reconciled with $b-\tau$ 
unification. On the other hand, 
for $\mu<0$,  $\Delta m_b$ is negative and the theoretical prediction 
for the $b$-quark mass can lie within the experimental range 
for values of $\tan\beta$ between roughly 30 and 40;
clearly, for a large $\tan\beta$ it is mandatory
to take into account
the large supersymmetric corrections to $m_b$ 
~\cite{Hall,CW,GNIS0}.  The figure also illustrates that, for $\mu>0$ 
and after taken properly into account supersymmetric corrections,
$m_b$ scales very slowly with $\tan\beta$. 
In other words, the renormalization flow for the bottom mass is 
attracted by some fixed value regardless of the initial 
conditions \cite{CW, FLL} (which as we see in the figure turns out
to be too large).
This {\it quasi-fixed} point, however, is not 
generated purely by the renormalization flow given by the RGE,
and  only appears after including supersymmetric corrections (the curve 
for $\Delta m_b = 0$ is not flat in Fig.~\ref{mb1}). Furthermore, 
for  $\mu<0$, there is no fixed point at all.

In the right panel of Fig. \ref{mb1} we repeat the 
analysis in the presence of massive neutrinos,
keeping only the third generation couplings (and ignoring
lepton mixing effects) from the $M_{GUT}$ to the
scale of the right-handed neutrino masses, $M_N$, 
and evolve the light neutrino
mass operator from this scale down to $M_Z$. 
A large value  of  the Dirac-type 
neutrino Yukawa coupling, $\lambda_N$, at the GUT scale
may arise naturally within the framework of Grand Unification, and
its value is determined, through the see-saw mechanism, 
by demanding a third generation
low energy neutrino mass of $m_{\nu_3}=0.05$~eV and evaluating the respective
heavy neutrino Majorana mass from the see-saw conditions. 
The predictions for $m_b(M_Z)$ using the 
lower and upper bounds of the 2-$\sigma$ experimental range of 
$\alpha_s$ and the corresponding range for $m_b(M_Z)$ after the evolution of the
bounds on $m_b(m_b)$ are shown for $M_N = 3\times 10^{14}$~GeV.
We use $m_t=172.6$ GeV \cite{mtop} in all computations. 

We observe that for $\mu>0$ the prediction 
for $m_b(M_Z)$ is always very large, despite its dependence on the 
soft terms through $\Delta m_b$. For $\mu<0$, there is a window 
of values of $\tan\beta$ compatible with asymptotic $b-\tau$ 
Yukawa unification. 
For the values of the soft terms considered in  Fig.{\ref{mb1}}, the allowed 
range of $\tan\beta$ moves from $27-44$ to $30-45$ when we introduce 
the effect of see-saw neutrinos. Let us stress that 
$b-\tau$ Yukawa unification is (is not) compatible with $m_b$, for 
$\mu<0 \left(>0\right)$, regardless on whether or not massive neutrinos 
are included, when lepton mixing effects are ignored.

\section{Yukawa Unification and $m_b$ for large lepton mixing}

The results are significantly modified once we consider the effects of lepton mixing
in the diagonalization and running of couplings from high to low energies. In order to
show this, we focus on $b-\tau$ 
unification within the framework of $SU(5)$ gauge unification and flavour symmetries that
provide consistent patterns for mass and mixing hierarchies, and naturally
reconcile a small $V_{CKM}$ mixing with a large charged 
lepton one. Taking into account the particle content of $SU(5)$ representations
(with symmetric up-type mass matrices, and down-type mass matrices that are transpose to
the ones for charged leptons), 
one finds that
\begin{equation}
{\mathcal {M}}_u \propto \left(\begin{array}{cc}
\bar{\varepsilon}^4 & \bar{\varepsilon}^2 \\
\bar{\varepsilon}^2 & 1 
\end{array}\right), \,\,\,\,
{\mathcal {M}}_d^0 \propto A\,\left(\begin{array}{cc}
0 & 0 \\
x & 1  
\end{array}\right), \,\,\,\,
{\mathcal {M}}_{\ell}^0 \propto A\,\left(\begin{array}{cc}
0 & x \\
0 & 1 
\end{array}\right)
\label{Mll}
\end{equation}
which, after diagonalization, lead to the relation
\begin{equation}
\frac{m_b^0}{1+x^2}=\frac{m_{\tau}^0}{1-x^2}\,\rightarrow \,
m_b^0=m_{\tau}^0\,\left(1-\underbrace{2x^2}_{\delta}+{\mathcal{O}}
\left(\delta ^2\right)\right)
\label{btau}
\end{equation}
where $\delta$ parametrizes  the flavour mixing in the (2,3) charged lepton
sector (any additional mixing required to match the data would then arise
from the neutrino sector). 

\begin{figure}[!h]
\hspace*{-0.3 cm}
\includegraphics[width=8cm,height=9cm]{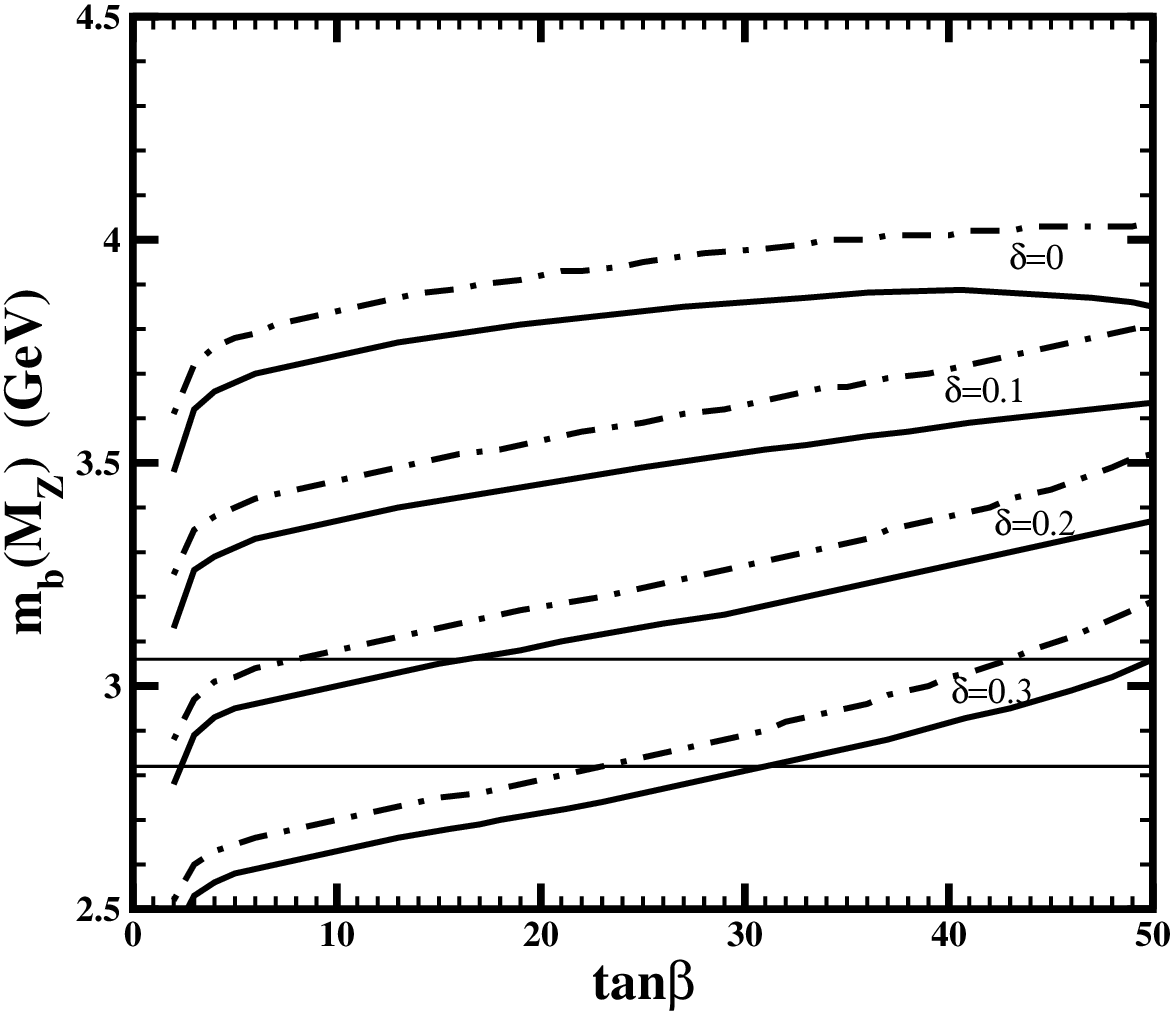}
\includegraphics[width=8cm,height=9cm]{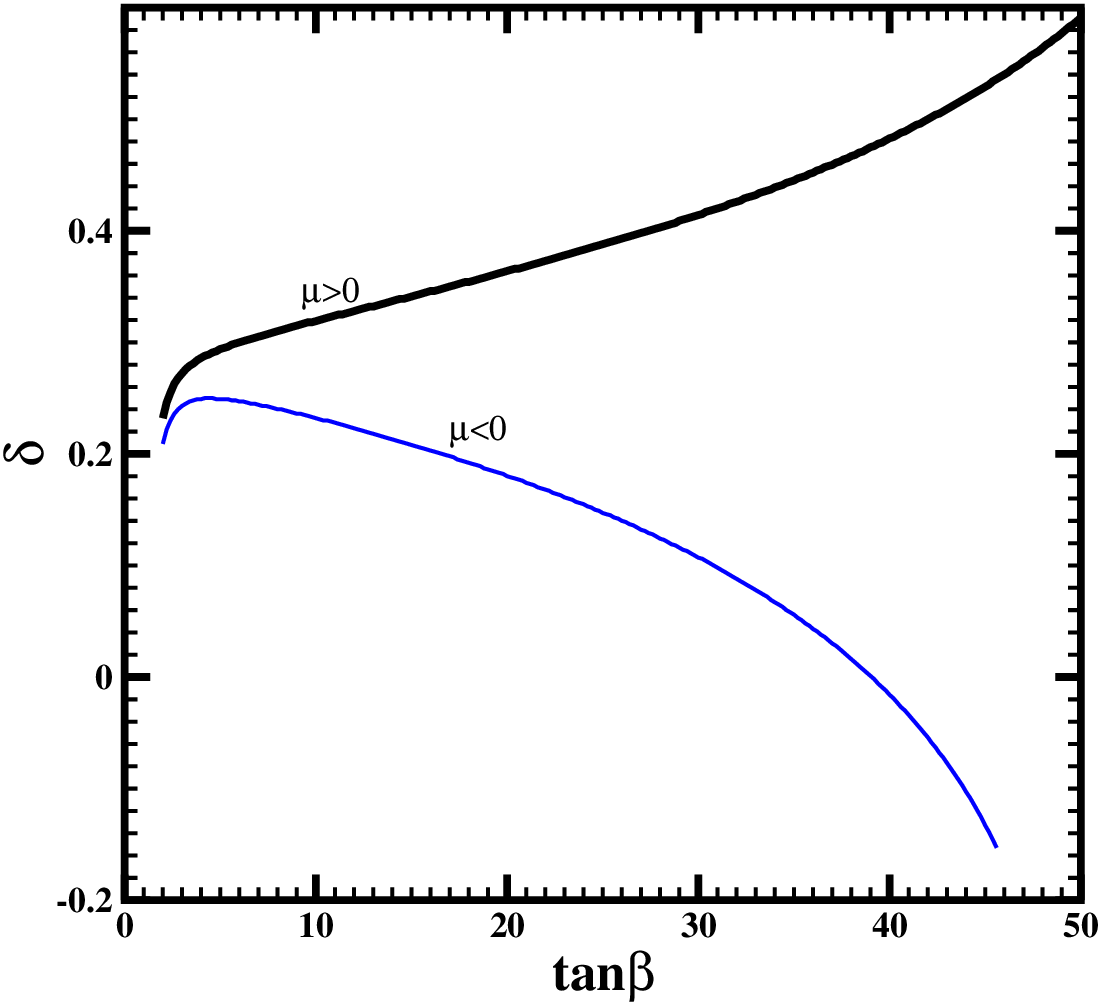}
\caption{\it 
In the left panel, we show $m_b$ as a function of $\tan\beta$ for $\mu>0$, 
without (solid lines) and with (dot dashed) massive neutrinos, in the latter case 
assuming $m_{\nu _3}=0.05$ eV and $M_N=3\times 10^{14}$ GeV. In the right panel, 
we show the required values of $\delta$ compatible with $b-\tau$ unification 
at the central experimental value of $m_b$ for both signs of $\mu$.
}
\label{deltb}
\end{figure}

In the left  panel of Fig.\ref{deltb} we show the change of $m_b$ as a 
function of $\tan\beta$, when the effects from large lepton mixing are correctly
considered. Comparing with the previous plots,  we see how solutions with 
positive $\mu$ are now viable, for the whole range of $\tan\beta$.
The appropriate size of the parameter $\delta$ in each case 
can be determined by imposing 
the relation $\lambda_\tau=\lambda_b(1+\delta)$ at $M_{GUT} $ and investigating
the values that are required in order to obtain a correct prediction for  
$m_b(M_Z)$. This is shown in the right panel of Fig.\ref{deltb}, where we 
demand a value of $m_b(M_Z)$ at
the center of its experimental range, for the central value of $\alpha_s$. 
We checked (although not shown) that the effect from runs above 
$M_{GUT}$ is very small, as expected. We then observe the following:
\begin{itemize}
\item
Solutions with a positive $\mu$ are now enabled, for the
whole range of $\tan\beta$.
\item
Negative $\mu$  is also  allowed in the whole range of $\tan\beta$ and requires 
a smaller mixing parameter $\delta$ than the $\mu>0$ case. 
\end{itemize}

\section{Yukawa unification and Dark Matter constraints}

In mSUGRA (or the CMSSM) for choices of the soft terms below the 
TeV scale, the LSP is mainly Bino-like and the prediction for 
$\Omega_\chi h^2$ is 
typically too large for models that satisfy the experimental constraints on 
SUSY. These constraints exclude models with relatively small values of 
$m_0$ and $M_{1/2}$ in which neutralino annihilates mainly through sfermion 
exchange. The values of WMAP can be obtained mainly in two regions:

\begin{itemize}
\item $\chi-\tilde{\tau}$ coannihilation region, which occurs for 
$m_\chi\sim m_{\tilde{\tau}}$.

\item  Resonances in the $\chi-\chi$ annihilation channel, which occur
for  $m_A\sim 2 m_\chi$.
\end{itemize}

\begin{figure}[!h]
\centering
\includegraphics[width=8cm,height=8cm]{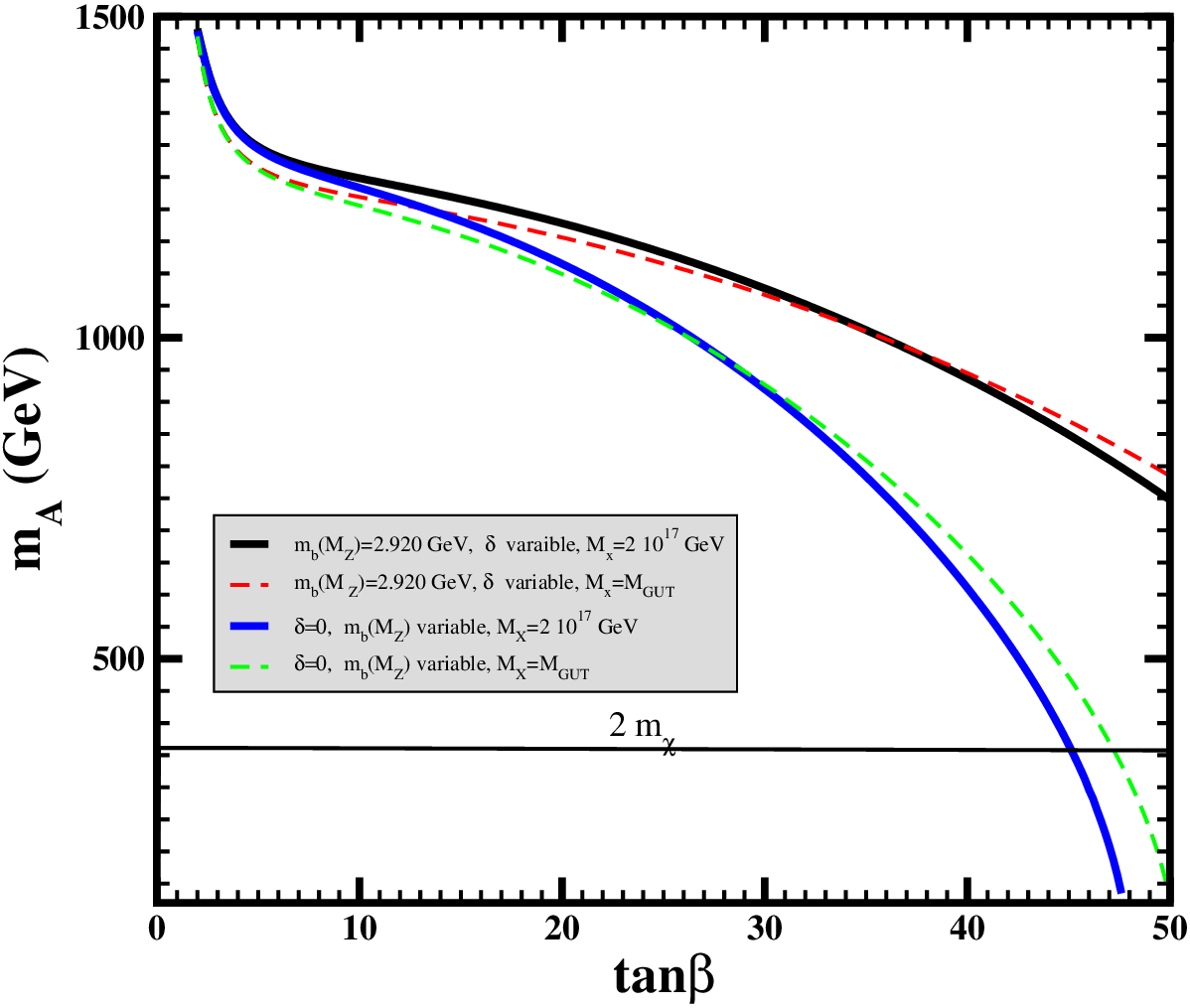}
\caption{\it Values of the pseudoscalar higgs mass $m_A$ versus $tan\beta$ 
for the 
same setting of parameters as in Fig.\ref{mb1}. In the upper lines 
the parameter $\delta$ varies so that  $m_b(M_Z)=2.921$~GeV, while 
in the lower lines $\delta=0$. The solid lines correspond to 
$M_X= 2\cdot 10^{17}$~GeV and the dashed lines to $M_X= M_{GUT}$.}
\label{matb}
\end{figure}

Since the above areas are tuned \cite{lahanas},  they will inevitably be 
sensitive to the changes induced by GUT unification and sizeable mixing in
the charged lepton sector. In this respect, 
it is illustrative to first investigate
the impact of the parameter $\delta$ 
 on the value of $m_A$. We show this in Fig. {\ref{matb}}, where 
we present the variation of 
$m_A$ with $\tan\beta$ for $m_0= 600$~GeV, $M_{1/2}=800$~GeV and 
$A_0=0$. In the upper lines the 
value of $\delta (\neq 0)$ is fixed by demanding $m_b(M_Z)=2.921$~GeV while 
in the lower ones $\delta=0$ with no restrictions on the  
$m_b(M_Z)$ prediction. We observe that the upper lines do not 
meet the resonance condition, while for
the lower lines this condition is achieved for values of 
$\tan\beta$ in the range $40-50$. 
We also see that, assuming soft term universality at a 
scale  $M_{X}>M_{GUT}$, the runs beyond $M_{GUT}$ change only moderately
the prediction for $m_A$.

Our next step is a global study of the supersymmetric
parameter space by assuming universal soft terms 
$m_0$, $M_{1/2}$ and  $A_0$  at some scale $M_X\geq M_{GUT}$, and obtain the 
mass spectrum by integrating the appropriate RGEs at each energy range,
as described above. The physical observables are computed 
using the code provided by {\it micromegas}~\cite{micro}. 
The 2-$\sigma$ range for BR($b \rightarrow s \gamma$) is constructed 
including an intrinsic $0.15 \cdot 10^{-4}$ MSSM  correction  as in Ref.~\cite{Ellis:2007fu}, leading to
\beq
2.15 \cdot 10^{-4}<BR(b \rightarrow s \gamma)< 4.9 \cdot 10^{-4}
\eeq

Without including the MSSM intrinsic error, the above range becomes
\beq
2.8 \cdot 10^{-4}<BR(b \rightarrow s \gamma)< 4.4\cdot 10^{-4}
\eeq
We take a conservative bound for the mass of the lightest $CP$-even Higgs of 
$m_h=114$~GeV and also consider  an uncertainty 
 of $\sim 3$~ GeV \cite{Heinemeyer:2004gx}
to its theoretical computation.

\begin{figure}[!h]
\centering
\includegraphics[width=7.5cm,height=8cm]{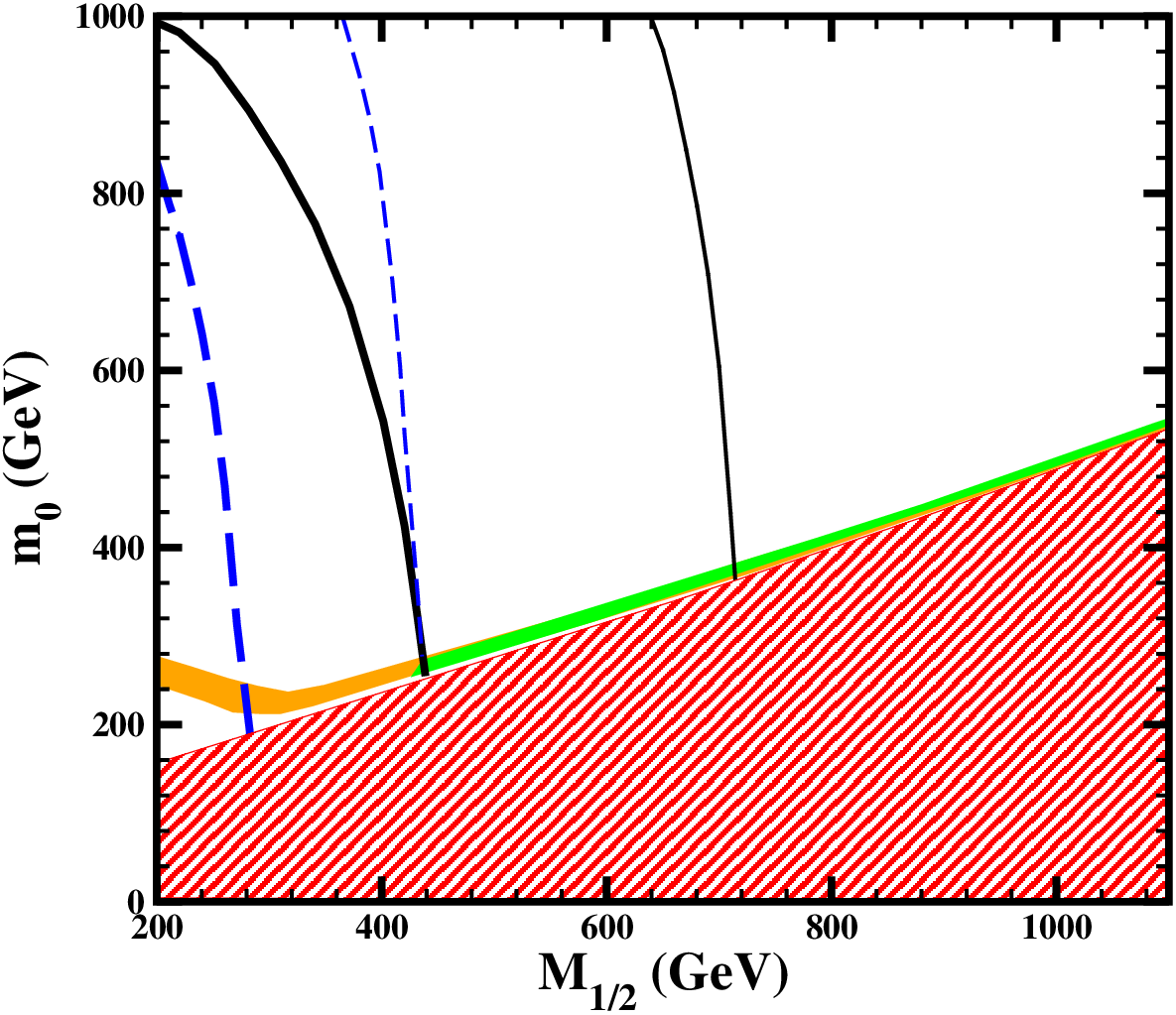}
\includegraphics[width=7.5cm,height=8cm]{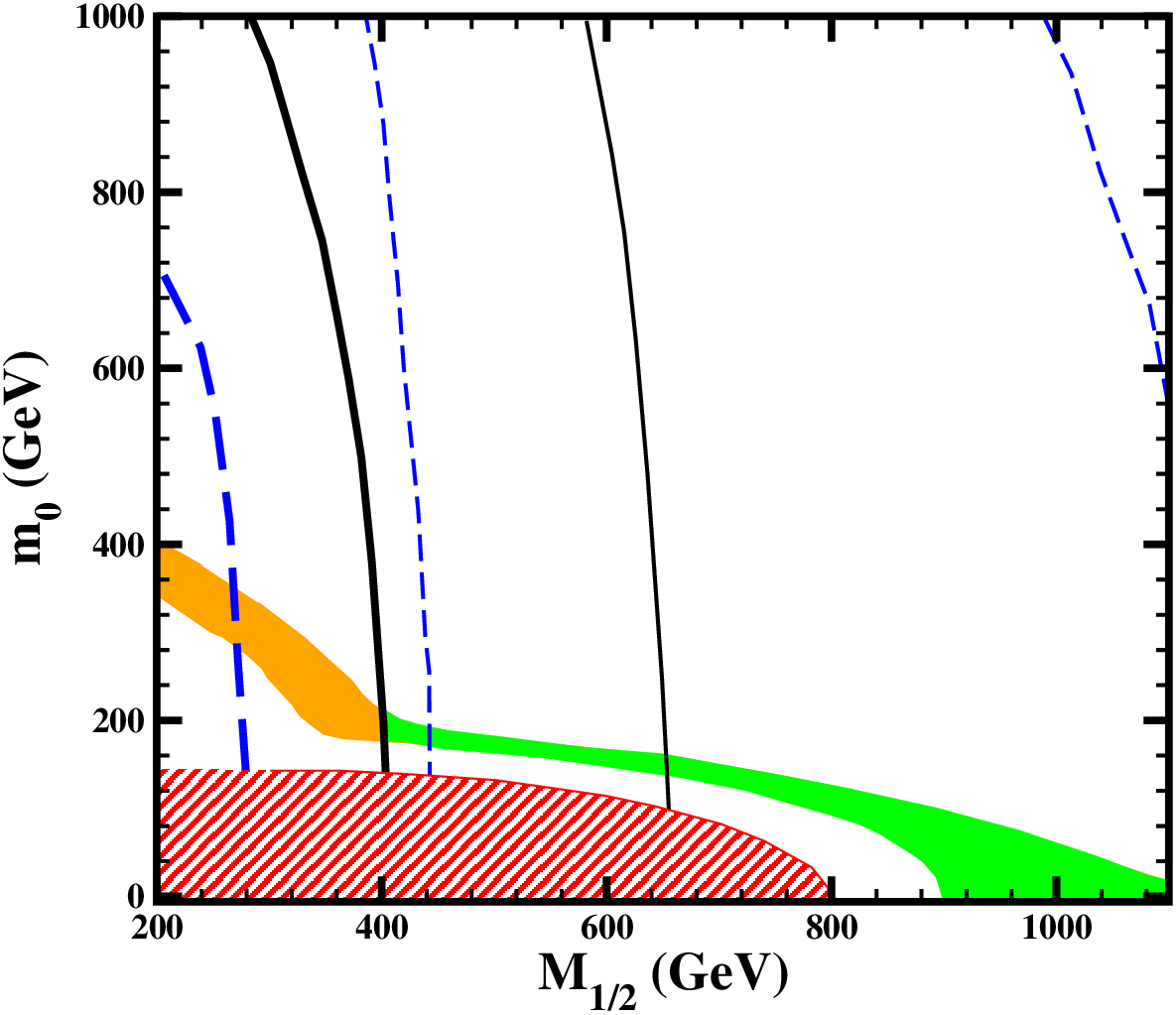}

\caption{\it  WMAP allowed area (green-shaded)
for the case of $\tan \beta=45$, $\mu>0$, $A_0=0$, 
when $\alpha_s(M_Z)$ and $m_b(M_Z)$ are set at their central values. 
The solid black lines indicates the $b \rightarrow s \gamma$ constraints 
[thick BR($b \rightarrow s \gamma)=2.15 \cdot 10^{-4}$, 
thin  BR($b \rightarrow s \gamma)=2.8 \cdot 10^{-4}$],  
and the dash blue line the Higgs mass bound [thin $m_h=114$~GeV, thick 
$m_h=111$~GeV].
On the left graph, $M_X=M_{GUT}$ while on the right panel,
$M_X=2 \cdot 10^{17}$ GeV.
}
\label{area45}
\end{figure}

In Fig.~\ref{area45} we present the WMAP favored area 
for $\tan\beta=45$, including lepton mixing effects
and considering the most tolerant bounds on 
 BR($b \rightarrow s \gamma)$ and  $m_h$ (stricter bounds on
 these constraints are also displayed).
In this example, we keep $m_b(M_Z)=2.920$~GeV which corresponds
to the evolution of the experimental value of $m_b$ obtained for
the central value $\alpha_s(M_Z)=0.1172$. When 
$M_{X}=M_{GUT}$ we find the WMAP favored region to be on the familiar CMSSM 
$\chi-\tau$ coannihilation area. However, as   in \cite{Mamb}, we find  
that the runs corresponding to 
$M_{X}>M_{GUT}$ have a big impact on the 
neutralino relic density and Fig. \ref{area45} shows clearly this effect.  The
  large values of the gauge unified coupling $\alpha_{SU(5)}$ tend to 
increase the values
of   $m_{\tilde{\tau}}$ in a way that, even if we start with $m_0=0$ at $M_X$ 
the model predicts $m_{\tilde{\tau}} > m_\chi$.
This implies that the allowed parameter space 
with a neutralino LSP is 
significantly enhanced (green area), while the coannihilation 
condition becomes harder to achieve; this
effect is more visible for large  $\tan\beta$ and  is 
understood since the lightest stau has a mass 
$m_{\tilde{\tau}_1}\simeq m_{\tilde{\tau}_{RR}}+ m_{\tilde{\tau}_{LR}}$, where
\begin{equation}
m_{\tilde{\tau}_{RR}}\simeq \left(1-\rho_\beta \right)m_0^2+0.3M_{1/2}^2, ~~
m_{\tilde{\tau}_{LR}}\simeq -m_{\tau}\mu\tan\beta
\end{equation}
$\rho_\beta$ being a positive coefficient dependent on $\tan\beta$ 
(in our case $\rho_\beta<1$) \cite{Mamb}. 
The increase of $m_{\tilde{\tau}_1}$ is due to 
the GUT runnings ($\sim M_{1/2}^2$). The picture for $tan\beta=35$ and $A_0=m_0$
 has been discussed in \cite{edson}, where it was shown that
the WMAP allowed region can be compatible with observable  flavour violation at the LHC.

\begin{figure}[!h]
\includegraphics[width=8cm,height=7cm]{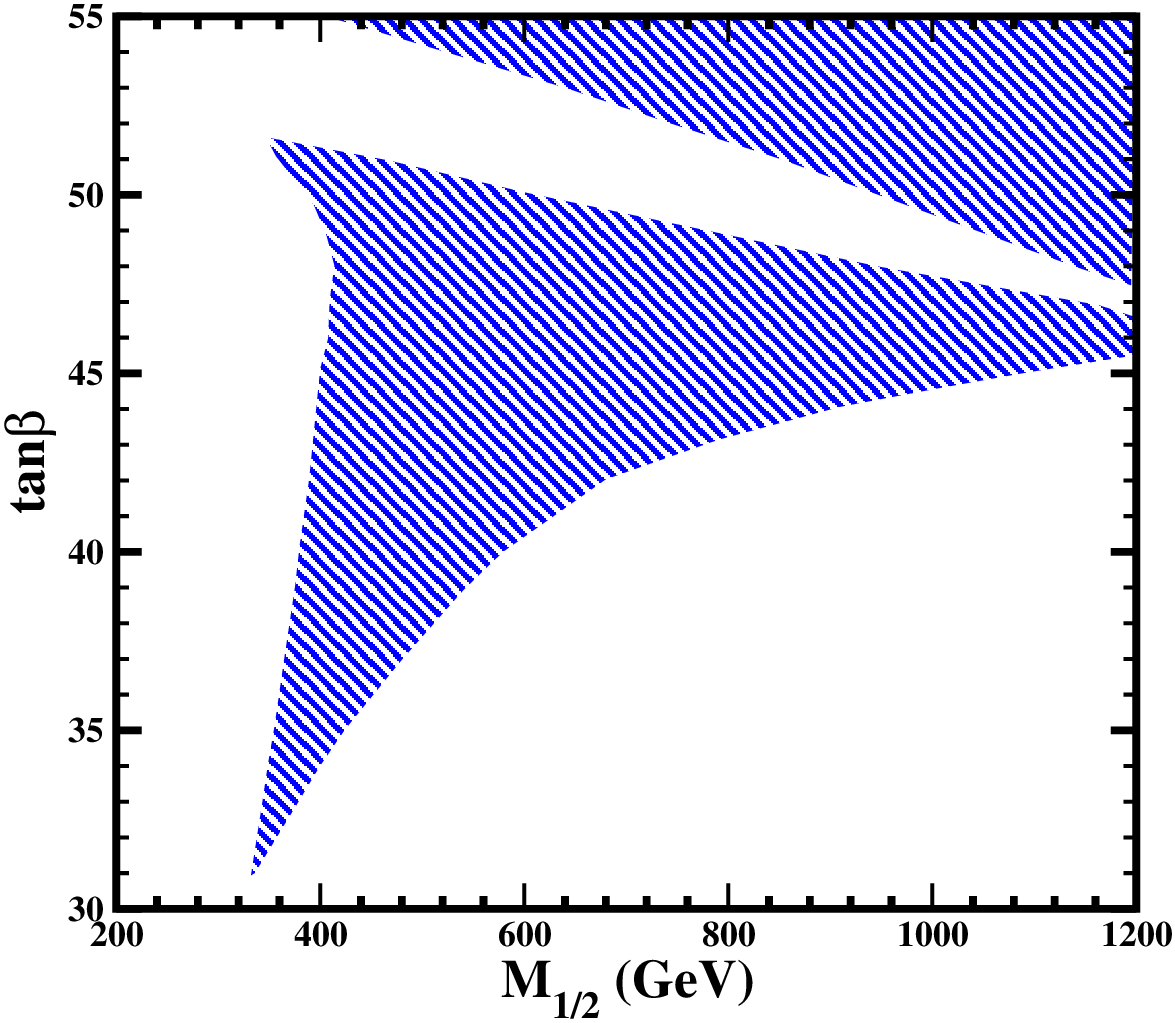}
\includegraphics[width=8cm,height=7cm]{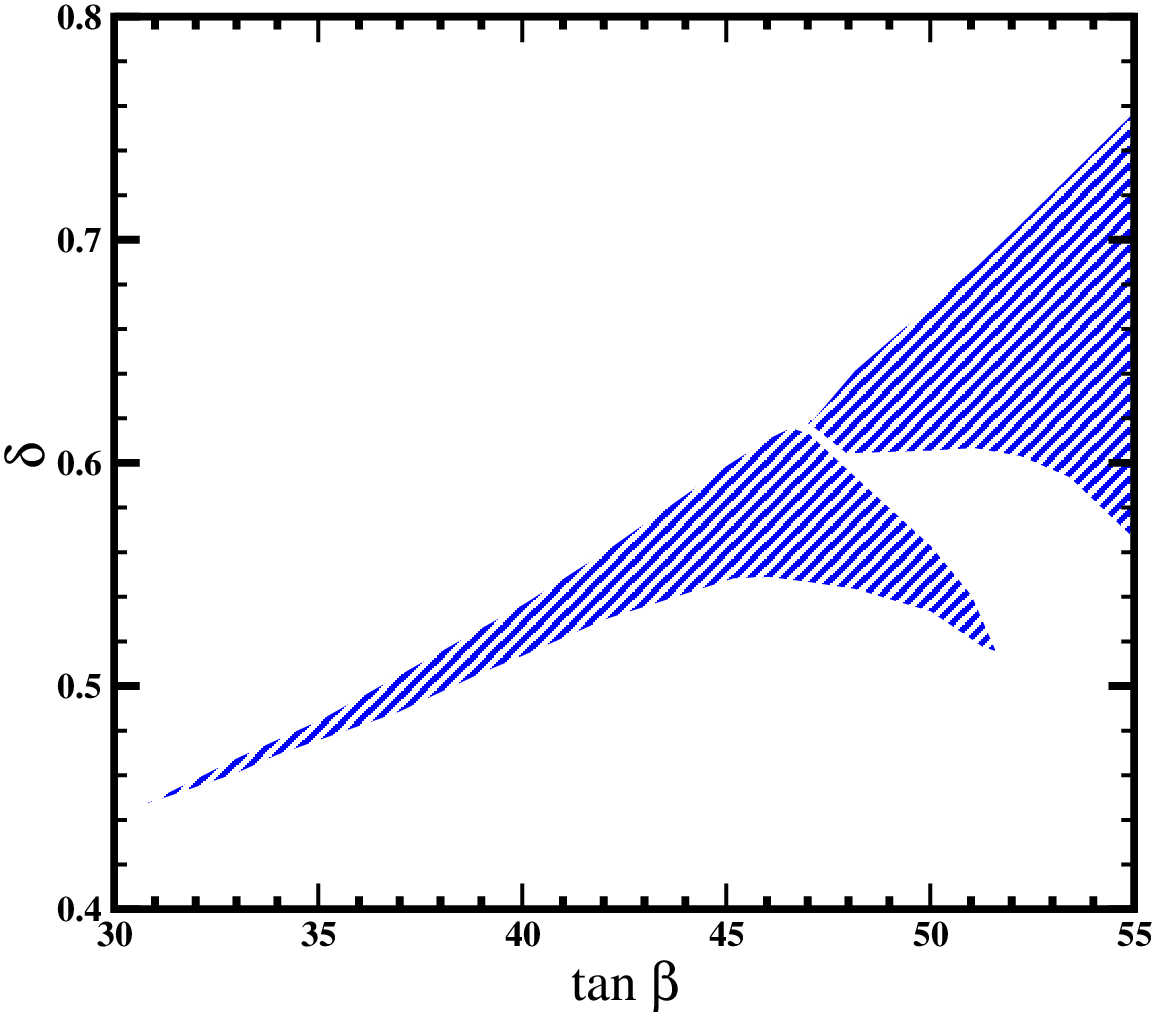}
\caption{\it 
The left panel provides a plot equivalent to 
Fig. 4 of \cite{Mamb} for the purpose of comparisons; 
the upper part corresponds
to areas with resonant Higgs channels. In 
the right panel, we show the values 
of $\delta$ corresponding to the previous plot; these inevitably implies 
$\delta\neq 0$.
}
\label{deltatb}
\end{figure}

\begin{figure}[!h]
\centering
\includegraphics[width=8cm,height=8cm]{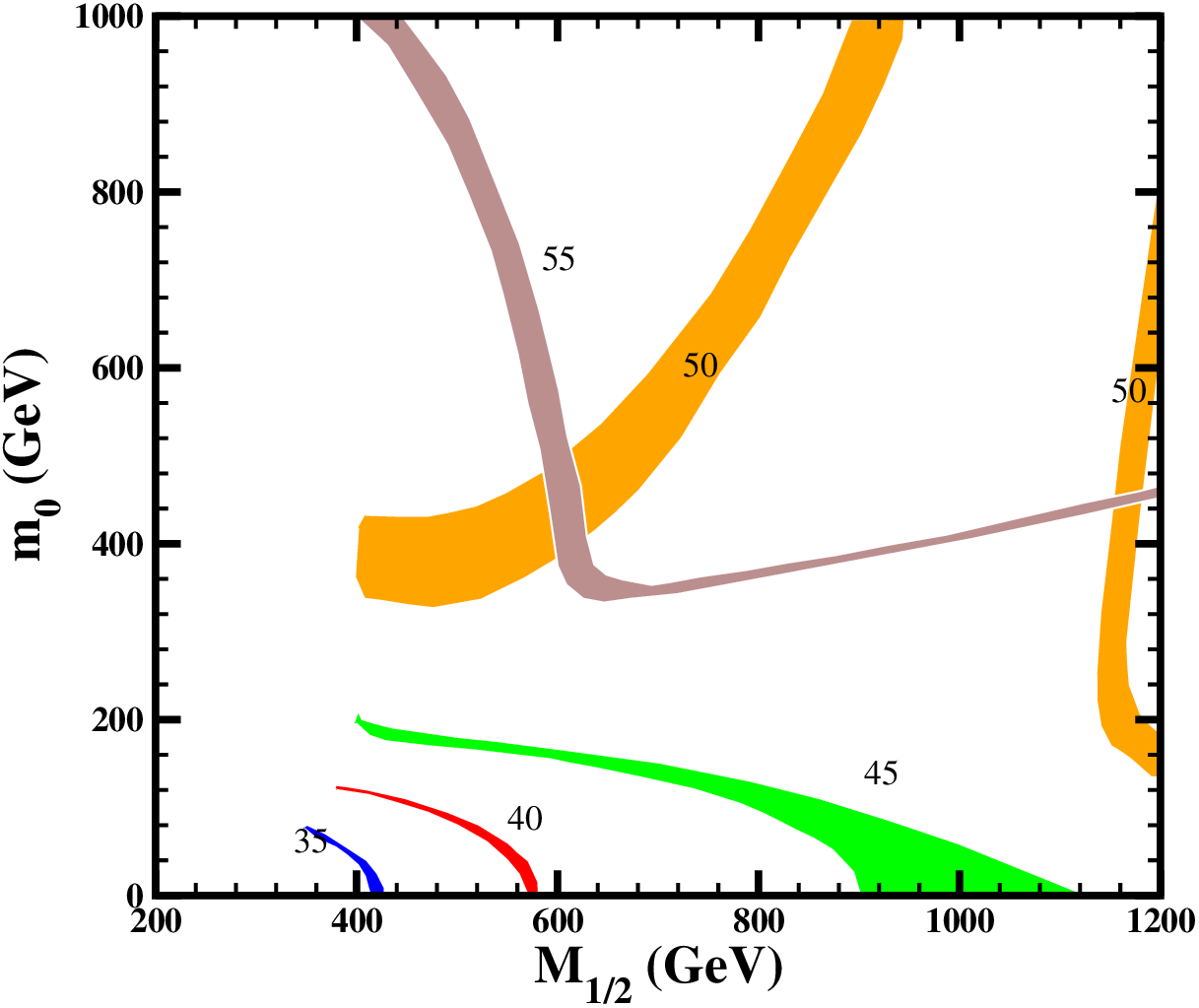}
\caption{\it
WMAP allowed areas for several values of $\tan\beta$. We consider 
$m_b(M_Z)=2.92$~GeV, $A_0=0$ GeV and $\mu>0$.
}
\label{m0m12}
\end{figure}

We stress again that $b-\tau$ Yukawa unification in the context of 
the CMSSM requires $\delta > 0$. An exhaustive  study of the parameter space 
for the case of 
$A_0=0$  and keeping the prediction of $m_b(M_Z)=2.92$~GeV is presented 
in Figs.~\ref{deltatb} and  \ref{m0m12}. The left panel of 
Fig.~\ref{deltatb}, shows 
that the phenomenological and cosmological constraints are fulfilled for 
$\tan\beta\ge 31$ and $M_{1/2}\ge 330$~GeV respectively. On the other hand, 
the right panel of Fig.~\ref{deltatb}
indicates that sizeable values of the mixing parameter $\delta$ 
are needed in order to maintain  $m_b(M_Z)$ in the experimental range 
and that the required mixing  increases with $\tan\beta$. 
Fig.~\ref{m0m12} shows the $\left(m_0,M_{1/2}\right)$ plane for different 
values of $\tan\beta$, taking all constraints into 
account; it can be seen that the resonant 
effects start becoming important at $\tan\beta\ge 45$. 

\begin{figure}[!h]
\centering
\includegraphics[width=7.5cm,height=8cm]{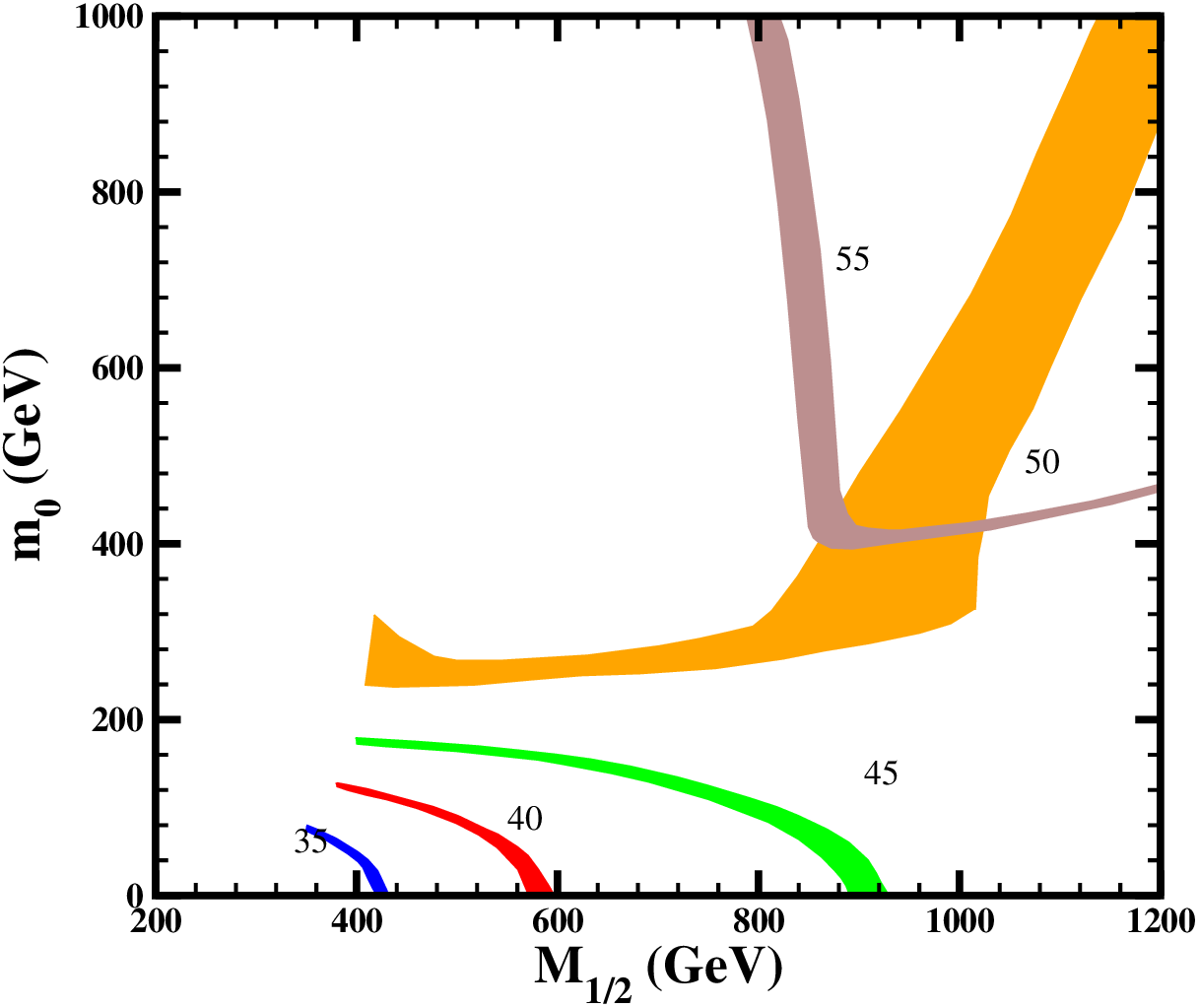}
\includegraphics[width=7.5cm,height=8cm]{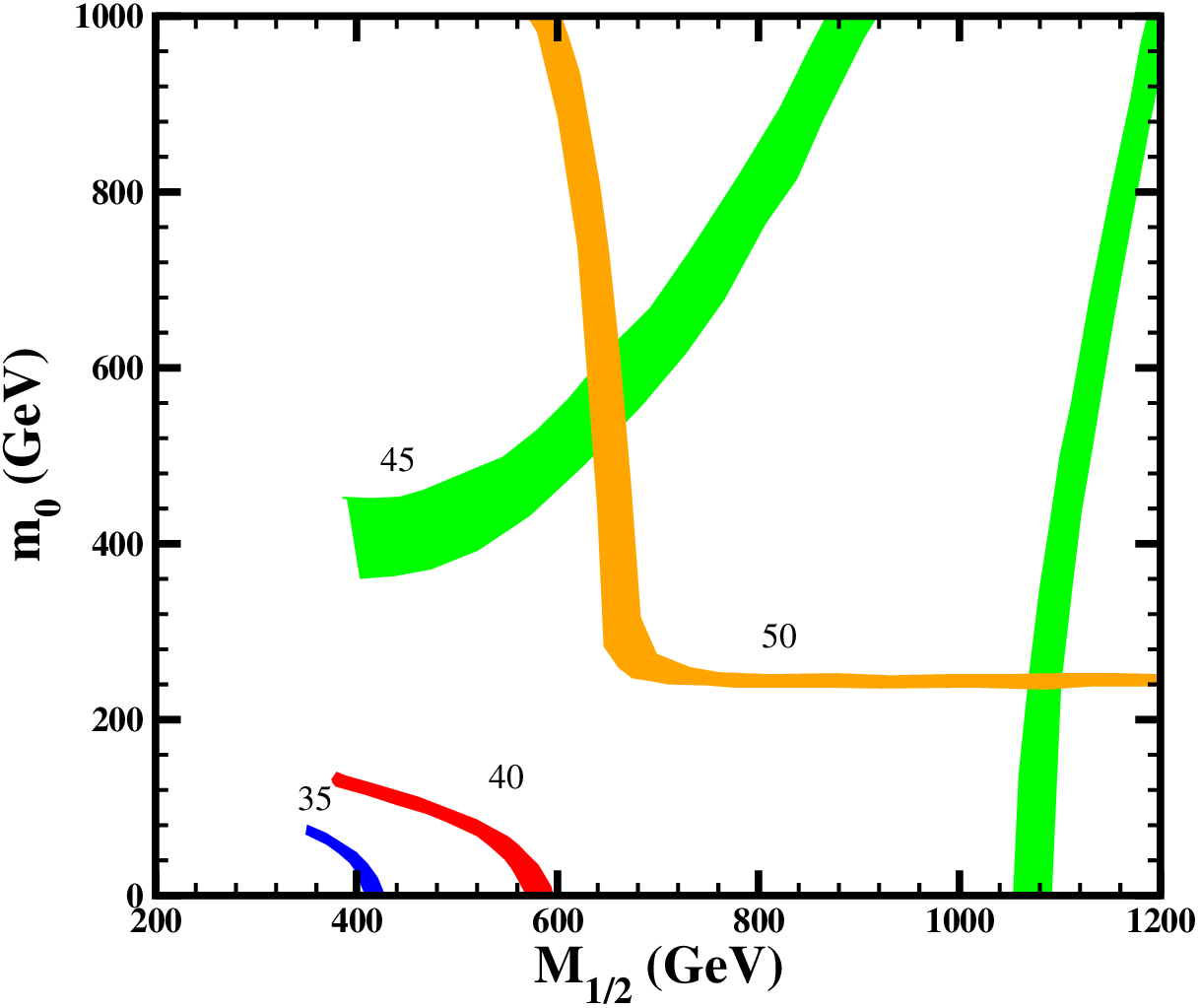}
\caption{\it Same plot as Fig.\ref{m0m12}, but taking the extreme values 
of both $\alpha_s(M_Z)$ and $m_b(M_Z)$. On the left panel, 
$\alpha_s(M_Z)=0.121$ and $m_b(M_Z)=2.74$~GeV while on the right panel 
$\alpha_s(M_Z)=0.113$ and  $m_b\left(M_Z\right)=3.114$~GeV. In both cases 
$M_X=2 \cdot 10^{17}$~GeV.
}
\label{m0m12a}
\end{figure}

The impact of varying $m_b$ and $\alpha_s$ within their experimental range 
can be seen in  Fig.~\ref{m0m12a} which indicates that,
 for  $\tan\beta$ below 40 there are no 
significant changes with the variation of $m_b$ on the WMAP area that lies  on the 
coannihilation region. For larger values of $\tan\beta$, however, small  
changes of the bottom Yukawa coupling due to the modification of $m_b$ have a 
significant impact on resonant annihilation, as is manifest by the locations 
of the bands for $\tan\beta=45$ and $50$ in the two lower plots.

The case $\mu<0$  is not yet ruled out by the $(g-2)_\mu$ data, 
since as we mentioned before by using $\tau$ data, a small 
negative discrepancy of the experimental 
measurement as compared to the SM prediction can be accommodated.  However, 
the upper limit on the BR($b \rightarrow s \gamma$) 
imposes a severe constraint on the SUSY parameter space 
since the supersymmetric contribution adds to the 
SM prediction. 

 \begin{figure}[!h]
\centering
\includegraphics[width=7.5cm,height=8cm]{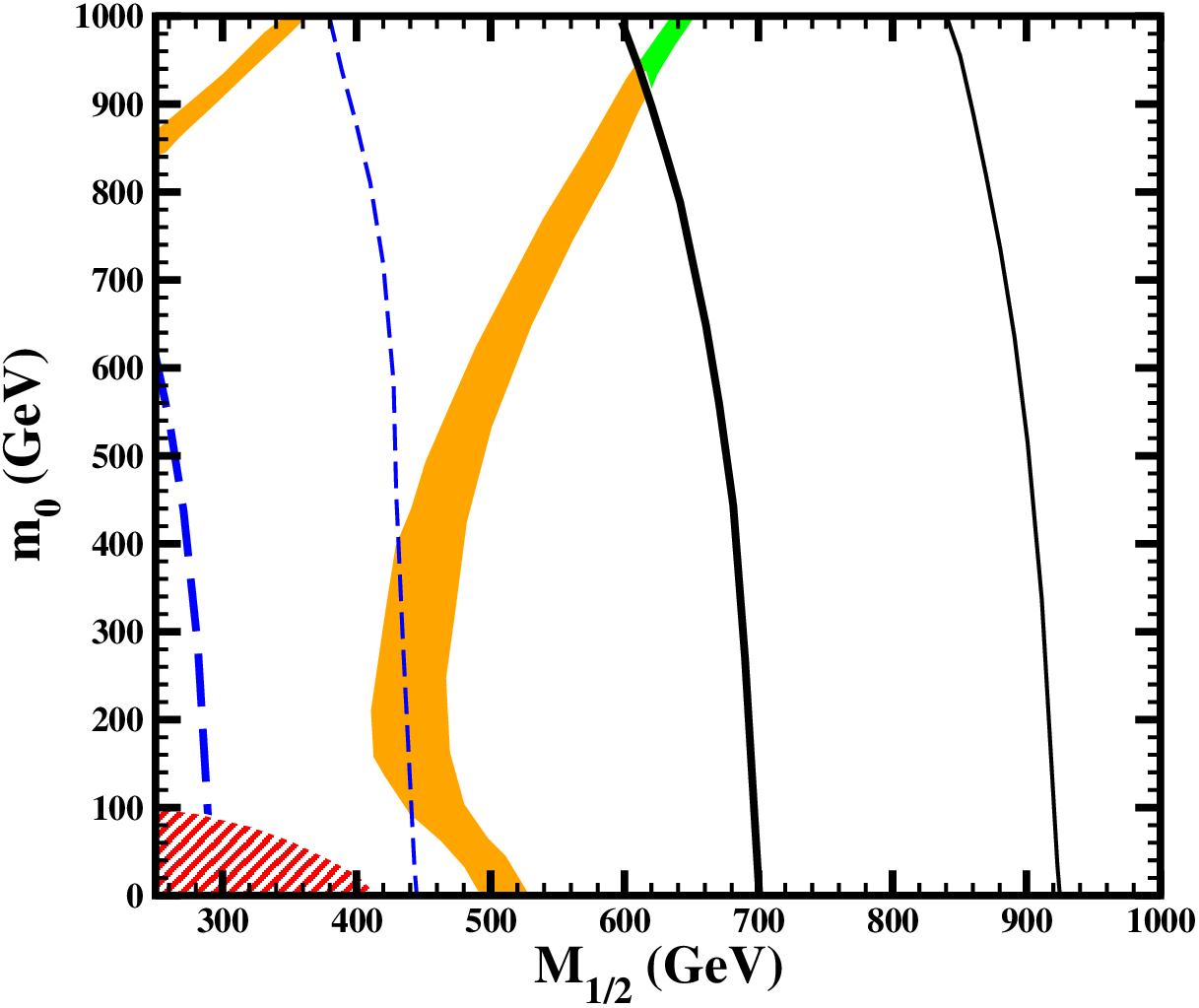}
\includegraphics[width=7.5cm,height=8cm]{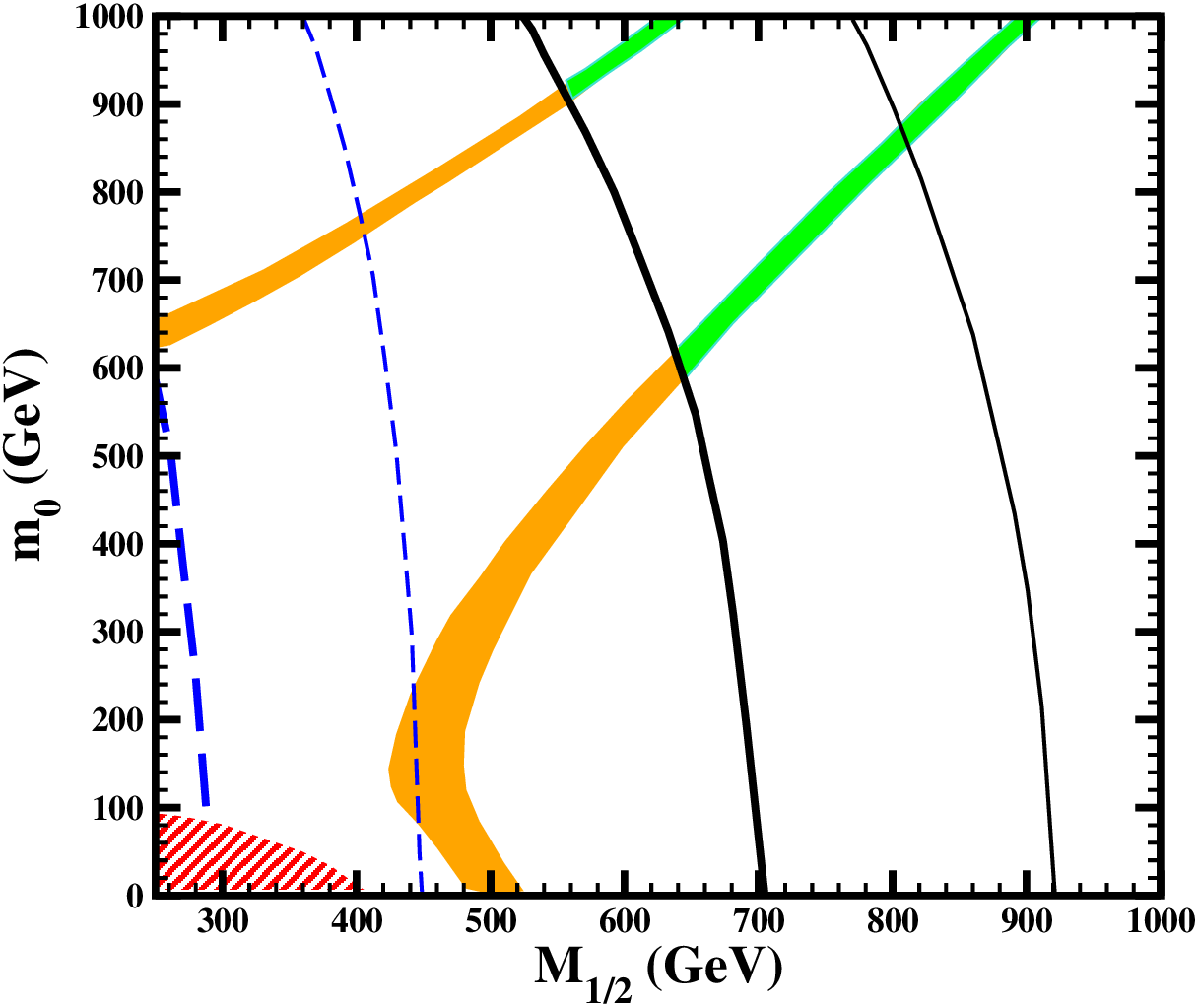}

\caption{\it $\left(m_0,M_{1/2}\right)$ plane for $tan\beta=40$, 
$A_0=0$, and $\mu<0$ . On the left panel we take the right-handed 
neutrino scale 
$M_N=6\times 10^{14}$~GeV, while on the right panel $M_N=10^{13}$ GeV. 
Here, $m_b(M_Z)$ 
and $\alpha_s(M_Z)$ are set to their central values. The lines follow the same 
notation as Fig.~\ref{area45}.
}
\label{m0m12n}
\end{figure}

For $\mu<0$ and $M_X>M_{GUT}$, we find that the WMAP areas allowed due 
to $\chi-\tilde{\tau}$ coannihilations at low values of $\tan\beta$ are excluded by 
the $b\to s \gamma$ higher bound. However, 
at larger values of $\tan\beta$, the areas with
resonant annihilations are marginally allowed, as we 
can see in Fig.~\ref{m0m12n}. 

As already underlined,
these results are very sensitive to $\lambda_N$. Once we 
decrease it by lowering $M_N$ (so that, from the see-saw condition,
$m_{\nu_3}$ remains 0.05~eV) we allow larger regions of the parameter space
for negative $\mu$ as well.  In both cases, however, the values 
of the parameter $\delta$ have to be very small ($\delta<0.05$); 
therefore, negative $\mu$ is not 
compatible with large charged lepton Yukawa mixing.

\section{Conclusions}

We revisited the WMAP dark matter constraints on Yukawa Unification 
in the presence of massive neutrinos. Large lepton mixing, as indicated 
by the data, modifies the predictions for the bottom quark mass,
and enables Yukawa unification also for large $\tan\beta$ and for 
positive values of $\mu$, which were  previously disfavoured. 
The larger the Dirac neutrino Yukawa couplings, the larger the effects.
A direct outcome is that the allowed parameter space
for neutralino  dark matter also increases, particularly 
when the effects of large lepton mixing are combined with runs
above $M_{GUT}$. Summarising, we find the following:

\begin{itemize}

\item $b-\tau$ unification is only allowed for $\mu>0$ in the presence of
large charged lepton mixing, which is motivated by the experimental data
of the recent years.

\item For  $\mu<0$ we still find values compatible with an exact 
$b-\tau$ unification at $M_X$. However, for large Dirac neutrino Yukawa couplings
and mixing arising dominantly from the charged lepton sector in the basis where the
down-quark mass matrix is diagonal, the space of parameters compatible with 
WMAP is i) only marginally compatible with the upper bound on 
$b\to s \gamma$ and ii) entirely excluded on the grounds 
of the $(g-2)_\mu$ observations, if we use $e^-e^+$ data to
estimate the SM vacuum polarization contribution.
  
\end{itemize}

Interestingly enough, it turns out that the cosmologically 
favoured parameter space also implies lepton flavour violating rates that are
very close to the current experimental bounds \cite{GLNR}.
Finally, additional interesting effects may arise in the case of non-universal
soft terms; these are also addressed in detail in \cite{GLNR}.

\vskip 1. cm
~\\
{\bf Acknowledgements} 
The authors would like to thank the European Network
of Theoretical Astroparticle Physics ENTApP ILIAS/N6 under contract number RII3-CT-2004-506222 for financial support. The research of S. Lola and P. Naranjo is funded by the FP6 
Marie Curie Excellence Grant MEXT-CT-2004-014297. The work of M.E.G and J.R.Q is supported by the Spanish MEC project FPA2006-13825 and
the project P07FQM02962 funded by âJunta de Andaluciaâ.

\vskip 1. cm

{\Large{\bf Appendix }}

In this appendix we summarize the RGEs that are most relevant for the 
purposes of the work addressed in this paper. For runs above the GUT scale 
the equations involving the Yukawa couplings and the soft mass terms 
corresponding to the $\mathbf{10}$ and $\mathbf{\bar{5}}$ representations 
of $SU(5)$, for the 3$^{rd}$ generation, take the form \cite{Hisano} 

\begin{equation}
16\pi ^2\,\frac{d\lambda _N}{dt}  = \left[-\frac{48}{5}g_5^2 + 7\lambda _N^2 
+ 3\lambda _t^2 + 4\lambda _b^2\right]\lambda _N\,\, ,
\label{A11}
\end{equation}

\begin{equation}
16\pi ^2\,\frac{d\lambda _d}{dt} =  \left[-\frac{84}{5}g_5^2 + 10\lambda _d^2 
+ 3\lambda _t^2 + \lambda _N^2\right]\lambda _d\,\, ,
\label{A12}
\end{equation}

\begin{equation}
16\pi ^2\,\frac{d\lambda _t}{dt}=\left[-\frac{96}{5}g_5^2 + 9\lambda _t^2 
+ 4\lambda _d^2 + \lambda _N^2\right]\lambda _t\,\, ,
\label{A13}
\end{equation}

\begin{eqnarray}
16\pi ^2\,\frac{dm_{\mathbf{10}}^2}{dt} & = & -\frac{144}{5}g_5^2\,M_5^2 
+ \left(12\lambda _t^2 + 4\lambda _d^2\right)m_{\mathbf{10}}^2 \\ \nonumber
 &  & + 4 \left[\left(m_{\mathbf{5}}^2 + m_{\bar{h}}^2\right)\lambda _d^2 
+ A_d^2\right] + 6\left(\lambda _t^2\,m_h^2 + A_t^2\right)\,\, ,
\label{A14}
\end{eqnarray}

\begin{eqnarray}
16\pi ^2\,\frac{dm_{\mathbf{5}}^2}{dt} & = & -\frac{96}{5}g_5^2\,M_5^2 
+ 2\left(4\lambda _d^2 + \lambda _N^2\right)m_{\mathbf{5}}^2 \\ \nonumber
 &  & + 8 \left[\left(m_{\mathbf{10}}^2 + m_{\bar{h}}^2\right)\lambda _d^2 
+ A_d^2\right] + 2\left(\lambda _N^2\,m_h^2 + \lambda _N^2\,m_{\mathbf{1}}^2 
+ A_N^2\right)
\label{A15}
\end{eqnarray}

For runs from $M_{GUT}$ to $M_N$, the equations for the Yukawa matrices are: 
\begin{equation}
16\pi^2\,\frac{d\lambda _N}{dt}=-\left[\left(\frac{3}{5}g_1^2 + 
3g_2^2\right)I_3 - \left(4\lambda _N^2+3\lambda _t^2+\lambda _{\tau}^2\right)
\right]\lambda _N\,\, ,
\label{A1}
\end{equation}

\begin{equation}
16\pi^2\,\frac{d\lambda _{\tau}}{dt}=-\left[\left(\frac{9}{5}g_1^2 + 
3g_2^2\right)I_3 - \left(4\lambda _{\tau}^2+3\lambda _b^2\lambda _N^2\right)
\right]\lambda _{\tau}\,\, ,
\label{A2}
\end{equation}

\begin{equation}
16\pi^2\,\frac{d\lambda _t}{dt}=-\left[\left(\frac{13}{5}g_1^2 + 
3g_2^2+\frac{16}{3}g_3^2\right)I_3 - \left(6\lambda _t^2+\lambda _b^2\right) 
+\lambda _N^2\right]\lambda _t
\label{A3}
\end{equation}

Since the neutrino has no coupling to the bottom quark, the Yukawa matrix 
corresponding to the latter remains unchanged with respect to the MSSM case. 

In section 2 semi-analytical expressions for the small $\tan\beta$ regime 
were given. In that case, only the top and the Dirac-type neutrino Yukawa 
couplings can be large at the GUT scale.

\end{document}